\documentclass[11pt, letterpaper]{article}
\usepackage[utf8]{inputenc}
\usepackage[margin=1.5cm]{geometry}
\usepackage{titlesec}
\usepackage{tabu}
\usepackage{enumitem}
\usepackage{amssymb}
\usepackage[dvipsnames]{xcolor}
\usepackage{graphicx}
\newlist{selectlist}{itemize}{2}
\setlist[selectlist]{label=$\square$,leftmargin=*,noitemsep,topsep=0pt}

\usepackage{lmodern}

\usepackage{hyperref}
\hypersetup{
    colorlinks=true,
    linkcolor=blue,
    filecolor=magenta,      
    urlcolor=blue,
}
 
\titleformat{\section}[block]{\hspace{1em}\bfseries}{\thesection.}{0.5em}{} 
\titleformat{\subsection}[block]{\hspace{1em}}{\thesubsection}{0.5em}{}

\begin{document}
\begin{flushleft}

\setlength{\parindent}{0pt}
\setlength{\parskip}{10pt}

\begin{center}
\Large
\textbf{Orthogonal-view Microscope\\ 
for the Biomechanics Investigations of Aquatic Organisms}\\
\end{center}

\begin{center}
\textbf{Authors}\\ 
 Brian T. Le $^{1,*}$; Katherine M. Auer $^{1,*}$; David A. Lopez $^{1,*}$; Justin P. Shum $^{1,*}$;  Brian Suarsana $^{1,*}$;\\
Ga-Young Kelly Suh $^{1}$; Per Niklas Hedde $^{2,\dag}$; and Siavash Ahrar $^{1,\dag}$. \\

(* Authors contributed equally; \dag Corresponding authors.)\\
(Version-6; Last updated: July. 24. 2023)
\end{center}

\textbf{Affiliations}\\ 
1. Department of Biomedical Engineering, 
California State University Long Beach\\ 1250 Bellflower Blvd. Long Beach, CA 90840, USA.\\
2. Beckman Laser Institute and Medical Clinic, University of California Irvine\\ Irvine, CA 92612, USA.\\
\hfill \break

\textbf{Corresponding contacts}\\  
E-mail: Siavash.Ahrar @ csulb.edu; Twitter: @SeeAvash\\
E-mail: Phedde @ uci.edu \\
\begin{center}
\textbf{Abstract}\\ \textit{ 
Microscopes are essential for the biomechanical and hydrodynamical investigation of small aquatic organisms. We report a do-it-yourself microscope (GLUBscope) that enables the visualization of organisms from two orthogonal imaging planes - top and side views. Compared to conventional imaging systems, this approach provides a comprehensive visualization strategy of organisms, which could have complex shapes and morphologies. The microscope was constructed by combining custom 3D-printed parts and off-the-shelf components. The system is designed for modularity and reconfigurability. Open-source design files and build instructions are provided in this report. Additionally, proof-of-use experiments (particularly with Hydra) and other organisms that combine the GLUBscope with an analysis pipeline were demonstrated to highlight the system’s utility. Beyond the applications demonstrated, the system can be used or modified for various imaging applications.}
\end{center}

\newpage
\textbf{Specifications table}\\

\vskip 0.2cm
\tabulinesep=1ex
\begin{tabu} to \linewidth {|X|X[3,l]|}
\hline  \textbf{Hardware name} & \textit{GLUBscope: the Orthogonal-view Microscope}
  \\
  \hline \textbf{Subject area} & %
 
  \vskip 0.1cm
  \begin{itemize}[noitemsep, topsep=0pt]
  \item \textit{Engineering and material science}
  \end{itemize}
  \\
  \hline \textbf{Hardware type} &
  \vskip 0.1cm  
  \begin{itemize}[noitemsep, topsep=0pt]
  \item \textit{Imaging tools}
  \end{itemize}
  \\ 
\hline \textbf{Closest commercial analog} &
   \begin{itemize}[noitemsep, topsep=0pt]
    \item \textit{No commercial analog is available}
    \end{itemize}
  \\
\hline \textbf{Open source license} &
  \begin{itemize} 
   \item \textit{CC-BY-SA 4.0}
  \end{itemize}
  \\
\hline \textbf{Cost of hardware} &
\begin{itemize} 
   \item \textit{ \$ 3000-4000}
  \end{itemize}
  \\
\hline \textbf{Source file repository} & 
  \textit{ \url{https://osf.io/9mw7d/}}
  \\

\\\hline
\end{tabu}
\end{flushleft}
\newpage
\section{Hardware in context}

Investigating biomechanics \cite{carter2016dynamics,stokkermans2022muscular,szymanski2019mapping}, hydrodynamics\cite{chan2010temperature}, and mechanobiology \cite{kim2018microfluidics} of organisms is of great interest. Beyond their fundamental biological importance, these investigations could provide approaches 
to mitigate or address the emerging threats to biodiversity due to climate change \cite{chan2012biomechanics}. Therefore, broader access to imaging tools, specifically microscopes, for laboratory and field investigations could be critical. Unfortunately, most commercially available microscopes are expensive, difficult to deploy for field studies, and challenging to customize \cite{nagalingam2023low}. Thankfully, microscopes have benefited from the emergence of the do-it-yourself (DIY) and open hardware movement. The DIY movement has reduced costs, provided access to local fabrication, allowed for customization, and increased access to imaging instruments \cite{wenzel2023open}. This report describes an open-sourced DIY orthogonal-view microscope for biomechanics and hydrodynamics investigation of aquatic organisms. \\

Over the past two decades, significant advancements have been made in developing low-cost microscopes and other DIY hardware. Salido et al. provide a comprehensive summary of and breakdown of these advances in their review \cite{salido2022review}. Broadly, in one approach to achieve cost-effective microscopy mobile phones are used in conjunction with DIY hardware and optics. Examples of this approach include a mobile phone strategy for clinical microscopy \cite{breslauer2009mobile}, a wide-field fluorescence microscope \cite{zhu2011cost}, Foldscope \cite{cybulski2014foldscope}, and CellScope \cite{tapley2013mobile}. Furthermore, the concept of mobile phone-based microscopy has expanded to techniques such as light-sheet \cite{hedde2021minispim}, enhanced micro and nanoparticle imaging \cite{sami2022excitation}, and other applications \cite{skolrood2022single}. A second approach combines DIY and open-source hardware to build microscopes from scratch, leveraging prototyping approaches (e.g., 3D printing and laser cutting). Del Rosario et al. provide key considerations for constructing microscopes using 3D printing \cite{del2022field}. Many exciting examples of DIY microscopes have been reported \cite{hohlbein2022open}. For example, the OpenFlexure project integrates optics with 3D-printed precision mechanical positioning, enabling automated microscopy that is accessible to a wide range of users \cite{collins2020robotic}. In another example, the UC2 (You. See. Too.) initiative has developed a modular and open-source toolbox that facilitates the construction of DIY microscopes \cite{diederich2020versatile}. Moreover, researchers have successfully produced 3D-printed components enabling imaging and control of model organisms (e.g., temperature) \cite{maia2017100}, as well as microscopes tailored for controlling optofluidic applications\cite{nagalingam2023low}. Of particular interest in the study of marine organisms are PlanktoScope \cite{pollina2023planktoscope,planktoscopeWeb} and the scale-free tracking microscope \cite{krishnamurthy2020scale,krishnamurthy2023active}. 
Combined with DIY microscopes, advances in fluidics have been critical in enabling biomechanics investigations of whole organisms (\emph{C. elegans}, zebrafish larvae, Drosophila larvae, and \emph{Hydra}) under well-controlled conditions \cite{kim2018microfluidics}. Specifically, microfluidics could immobilize, trap, or limit the organisms' movements to enable imaging or other functional recordings \cite{wong2013live,san2018microfluidics}. These strategies are critical for many investigations due to organisms' propensity for shifting out of focus.  However, this approach may limit some biomechanics considerations. However, maintaining organisms (or key features) in the field of view can be challenging due to their shape or small rotations. \\ 

In this investigation, we developed an imaging system for the biomechanics investigation of aquatic organisms. The DIY microscope, GLUBscope, was built by combining custom 3D-printed parts and broadly available off-the-shelf components. One key feature of GLUBscope is the ability to view the samples from two orthogonal planes of imaging - a top view and a side view. With this approach, an organism could move out of a plane in one view yet remain visible in the orthogonal view. Due to their complex 3D shapes, viewing organisms from multiple angles could be critical in addressing questions related to biomechanics and hydrodynamics (e.g., surface attachment, elongation, or stretching in multiple planes). In this effort, we avoided the immobilization strategies and used chambers (3.5 mL volume) with larger volumes than the standard microfluidics often used for biomechanics investigations. The system was built for modularity and reconfigurability, such that various components could be readily exchanged. In this report and accompanying repository, we have shared design files, bill of materials, and build and operation instructions. Additionally, proof-of-use experiments with organisms (particularly \emph{Hydra}) are provided to demonstrate the use of GLUBscope for biomechanics investigation of aquatic organisms. 

\section{Hardware description}
GLUBscope provides two orthogonal views of the samples, setting it apart from conventional imaging systems. To build the GLUBscope, off-the-shelf components, and DIY 3D-printed parts were used. The system was built on an optical breadboard measuring 12" x 18" x $1/2$". We note that almost all commercial mechanical components (e.g., optical breadboard) can be replaced with alternatives. They were used here due to their broad accessibility.\\

To position samples, an XYZ micromanipulator stage was used. \textbf{Figure 1} presents images of the GLUBscope and a simplified system block diagram emphasizing the optical components. The 3D-printed parts were made via Snapmaker 2.0. These parts included two guide rails to move optical components to focus the image. Two cameras were used, one for capturing the top view (or vertical view) and another for the side view (or horizontal view). We used FLIR cameras and Arducam cameras, offering a choice between high-resolution/high-sensitivity and more cost-effective options. Camera control was achieved using Micromanager software and Arducam software. The study used various objectives ranging from 4X to 10X to suit different magnification requirements. Brightfield imaging of the samples was achieved by employing two bright white LEDs positioned below and on the side of the sample. For horizontal (side) illumination, a stand was developed. An optional diffuser holder was developed for top view illumination. LEDs were powered by a 9V battery and regulated using two resistors (variable or fixed at 350 $\Omega$) such that the system could be operated in the field without access to an electrical grid. The components were secured onto the optical breadboard via 1/4-inch 20 screws. The bill of materials, the designs for the 3D-printed parts, and other resources are provided through an Open Science Framework (OSF) page.  \linebreak

Potential uses of the GLUBscope include:
\begin{itemize}[noitemsep, topsep=0pt]
    \item \textit{Orthogonal view imaging (top and side).}
    \item \textit{Characterize biomechanics or hydrodynamics of aquatic organisms.}
    \item \textit{Customize and use as a field microscope.}
    \item \textit{Use for pedagogy or classroom room setting.}
    \end{itemize}

\section{Design files summary}
Editable versions of design files for the 3D-printed components are available from the project's Open Science Framework repository, {\url{https://osf.io/9mw7d/}}. 

Following is the summary of files for the hardware. \\
\vskip 0.1cm
\tabulinesep=1ex
\begin{tabu} to \linewidth {|X[2,1]|X|X|X[1.5,1]|} 
\hline
\textbf{Design Filename} & \textbf{File Type} & \textbf{License} & \textbf{Location of the File} \\\hline

breadboard$\_$and$\_$battery$\_$holder  & 3D Model (.stl) & CC-BY-SA 4.0 & OSF page \\\hline
horizontal$\_$camera$\_$holder & 3D Model (.stl) & CC-BY-SA 4.0 & OSF page \\\hline
horizontal$\_$camera$\_$path & 3D Model (.stl) & CC-BY-SA 4.0 & OSF page \\\hline
horizontal$\_$camera$\_$post & 3D Model (.stl) & CC-BY-SA 4.0 & OSF page \\\hline
horizontal$\_$camera$\_$stands & 3D Model (.stl) & CC-BY-SA 4.0 & OSF page \\\hline
horizontal$\_$light & 3D Model (.stl) & CC-BY-SA 4.0 & OSF page \\\hline
vertical$\_$camera$\_$adapter  & 3D Model (.stl) & CC-BY-SA 4.0 & OSF page \\\hline
vertical$\_$camera$\_$holder & 3D Model (.stl) & CC-BY-SA 4.0 & OSF page \\\hline
vertical$\_$camera$\_$posts & 3D Model (.stl) & CC-BY-SA 4.0 & OSF page \\\hline
vertical$\_$camera$\_$stands & 3D Model (.stl) & CC-BY-SA 4.0 & OSF page \\\hline
diffuser holder v2 & 3D Model (.stl) & CC-BY-SA 4.0 & OSF page \\\hline 
\end{tabu}

\vskip 0.3cm
\noindent
The supplementary material provides an animated rendering of the components' assembly.

\section{Bill of materials}
A complete bill of materials is available from the OSF page with additional comments. The estimated costs and potential vendors are provided. We note that availability and costs are subject to suppliers. Alternative options are recommended when possible. 

\section{Build instructions}
Step-by-step instructions for the system build, and images are provided (\textbf{Figures 2 and 3}). Please note that, for the connections listed below relying on $1/4$” - 20 screws, it is possible to reverse the orientations while constructing the GLUBscope with no negative impact on build stability. Four screws are the minimum requirement for a secure connection between the base of a part and the optical breadboard. There are additional ports available for a more secure attachment. A degree-marked level (circular bubble level bullseye) can be used to ensure that the parts are level.\\

\begin{enumerate}
  \item[$\blacksquare$] SAFETY 1:  Please use proper personal protective equipment (PPE). Bright LEDs and other light sources could pose a safety risk.
  \item[$\blacksquare$] SAFETY 2: Recall that the bead of a 3D Printer can become hot. After the print is completed, remove the part from the bed by applying a gentle shear force. If a chisel is required to remove the part, please use appropriate PPE, such as cut-proof gloves and safety glasses.
  \item[$\blacksquare$] SAFETY 3: Please use appropriate ventilation and PPE when soldering. 
  
\end{enumerate}

Following are the instructions to build the GLUBscope.

\begin{enumerate}

\item[] \textbf{Horizontal Path Assembly}

\begin{enumerate}

\item  Place the M6 breadboard on a flat surface. Here the position (port) on the top left corner is referenced as coordinate $(1,1)$.

\item Optional Step: Apply double-sided (set) screws in the 4 corners $(1,1),  (1,12),  (18,1),  (18,12)$. Attach optical construction rails to create an enclosure for the system to block ambient light. 

\item Roughly align the parts \textbf{vertical\_camera\_stands} in desired coordinates $(9,1),(9,10)$, spaced 166 mm apart or with 6 ports spaced between them.

\item Secure the stands via four screws. 

\item Align \textbf{horizontal\_camera\_stands} centers at coordinates $(3,7),(8,7)$, spaced 58.44 mm outside to inside. 

\item Fasten screws into 4 corners. See \textbf{Figure 2 A}.

\item Place \textbf{horizontal\_camera\_path} on top of \textbf{horizontal\_camera\_stands} and secure in place with screws. See \textbf{Figure 2 B}.

\item Slide \textbf{horizontal\_camera\_posts} into \textbf{horizontal\_camera\_path}. See \textbf{Figure 2 D and E}.

\item Attach FLIR camera with optical components (See \textbf{Figure 2 E} attached to \textbf{horizontal\_camera\_holder}.

\item Slide \textbf{horizontal\_camera\_holder} with camera attached to \textbf{horizontal\_camera\_post}.

\item Connect FLIR camera to USB 3.1 Gen 1 Micro-B to USB-A Locking Cable and then the computer. See \textbf {Figure 2 H}.

\end{enumerate}

\item[] \textbf{Vertical Path Assembly}

\begin{enumerate}
\item Align \textbf{vertical\_camera\_posts} with \textbf{vertical\_camera\_stands} and secure in place with screws.

\item Slide \textbf{vertical\_camera\_holder} on \textbf{vertical\_camera\_posts}, with the short side closest to the user.

\item Place \textbf{vertical\_camera\_adapter}  on the top of \textbf{vertical\_camera\_holder} and align ports. See \textbf{Figure 2 I and J}.

\item Align FLIR camera in \textbf{vertical\_camera\_adapter\_flir}. See \textbf{Figure 2 K}.

\item Attach the optical path with desired lens, filter, and objective on the bottom port of \textbf{vertical\_camera\_holder}
Connect FLIR camera to a computer via USB 3.1 Gen 1 Micro-B to USB-A Locking Cable. See \textbf{Figure 2 L}.

\item Alternatively, the Arducam camera can be used. To connect the camera to the computer a USB USB2.0 cable were used.
\end{enumerate}

\item[] \textbf{Lighting System Assembly}

\begin{enumerate}
\item Solder jumper wires to the 3W LEDs (flat form factor) for vertical illumination. Please note one or two LEDs may be needed. LEDs with a standard form factor can be used for the side illumination. These LEDs can be directly attached via jumper wires. 

\item Attach the 3W LED on the optical breadboard. Roughly align the LED with the objective for vertical imaging between $(12,6)$ and $(12,7)$. See \textbf{Figure 3 A}.

\item Run wires underneath optical breadboard in port $(13,6)$.  The electrical breadboard connects these wires to the circuit and 9V battery.  

\item Use foam materials inside the rectangular opening of the \textbf{horizontal\_light} for securing the side illumination LED. Then, attach \textbf{horizontal\_light} to the optical breadboard via $(14,5), (14,6)$. \textbf{Figure 3 B}.

\item Connect the LEDs to jumper cables. Roughly align the LED with the horizontal objective. 

\item Move the wires underneath optical breadboard via $(15,6)$ to be later connected to the electrical circuit.

\item Attach the \textbf{breadboard\_and\_battery\_holder} to optical breadboard via $(3,2), (3,3)$ and $(7,2),(7,3)$ positions.

\item Attach the Micromanipulator stage into ports $(11,8)$ and $(13,8)$. 

\item Place the electrical breadboard and a 9V battery in their respective holders.

\item Using two 350 $\Omega$ resistors (or two potentiometers) and necessary wires, create two series circuits to power the LEDs via the 9V battery. See \textbf{Figure 3 F}. Please note that it is important to select resistors (or two potentiometers) with appropriate power rating. 

\item Bring the wires from both LEDs to above the optical breadboard through ports $(8,3)$ and $(8,2)$ and then connect them to the electrical circuit. See \textbf{Figure 3 F}.

\item Place the \textbf{diffuser holder v2} above the 3W LED for vertical illumination. See \textbf{Figure 3 D and E} for the position.

\item	Place an optical diffuser inside of the \textbf{diffuser holder v2} for better illumination.
\end{enumerate}
\end{enumerate}

\section{Operation instructions}

\subsection{Camera configurations:} The GLUBscope cameras can be operated with one or two computers. In this, we typically used two computers. The use of more high resolution (20 MP) and economical cameras was demonstrated. Two 20 MP cameras (FLIR) were used for the first option. These were connected via a USB 3.1 Gen 1 Micro-B to USB-A Locking cable and controlled via open-source Micromanager software.
Please note that it is important to select the correct hardware configuration files.  
The OSF repository provides guides (written and video instructions) for the configuration selection. For the economical option, Arducam cameras (webcams) were used. These cameras were connected to the computer with a USB 2.0 cable . They were controlled via the AMCap software. \\

\subsection{Sample positioning:} All components should be level for the best alignment. Next, turn on the LEDs to confirm that the cameras can collect live images. The rail guides are used for rough focus, and the XYZ micromanipulator stage is used to fine-focus the sample. Typically an edge of a slide or a cuvette can be used as a target for focusing an image. Please note that the light sources can be moved to better illuminate the entire field of view. In our experience, the LEDs could be too close to the sample, only illuminating a portion of the field. The optional diffusers (for the top view) improved the illumination. Please note that simultaneously viewing the same field of view (top and side) may require adjustments. Using a lower magnification for the top view was helpful to this aim. Our investigations used a 4X objective for the top view and a 10X for the side view. The working distance of the objectives was not an issue. This parameter should be considered for higher magnification objectives. Objectives can be changed based on a project's needs. Additionally, LEDs can be adjusted - the distance between the light source and the sample and brightness - to better illuminate the sample. 

\subsection{3D printing:} Components were 3D printed via a 
Snapmaker 2.0 with Polylactic Acid (PLA) filaments. The final version of guide rails was 3D printed with nylon composites to ensure mechanical durability.  

\section{Validation and case studies}

GLUBscope was validated and proof-of-use experiments were investigated with the following experiments. 

\textbf{Calibration slides}: Standard slides (i.e., a 1951 Resolution Test Target slide and a 0.01mm calibration slide) were used to obtain top and side images. Images generated from the references are presented in (\textbf{Figure 4 A and B}). Please note that the top view of 0.01mm is illuminated via an LED without a diffuser. GLUBscope was also used as a fluorescence microscope. To this aim, the white LEDs were switched with bright green LEDs. 5 $\mu$m green fluorescent beads were used as the sample. Beads were suspended in DI water and were added to the cuvettes (3.5 mL volume, 10mm path length, 4 Clear Windows). Preliminary results are presented as supplementary material (\textbf{SupFigure 1}). We note this feature requires further optimizations to improve the excitation light source. 

\textbf{Organism Biomechanics}: Proof-of-use experiments were conducted to demonstrate using GLUBscope in biomechanics investigations. In the first demonstration, freshwater polyps \emph{Hydra vulgaris} were imaged using the top-view (4X) and side-view (10x) (see \textbf{Figure 5A}). A simple analysis pipeline was also developed to enable biomechanical investigations\textbf{Figure 5B}. Using the video recordings from the microscope and DeepLabCut application \cite{nath2019using}, anatomical landmarks from the organism were tracked. These included features such as the foot, two points across the body column, tentacles, and the head \textbf{Figure 5B}. In the presented position plot, blue squares represent the starting position, and yellow triangles mark the final position. Changes in position are highlighted by lines connecting these two shapes with color coding encoding time. Moreover, two simple applications, \emph{Frame to segment} and \emph{Segment to distance}, were developed for additional biomechanical analysis. Using the position data generated by DeepLabCut, initial and final length segments and percentage elongation of the segments were calculated.
The source code for both applications is available from the OSF repository. \\

Next GLUBscope was used to visualize sand dollar larvae (\emph{Dendraster Excentricus}) and sea anemones \textbf{SupFigure 2}. In the case of larvae, GLUBscope enabled the imaging of the complex structure of the larvae from multiple angles. While preliminary, this ability could provide a unique approach to visualizing the complex hydrodynamics of these organisms. Side visualization was critical for the final demonstration, imaging sea anemones since the organisms were attached to an opaque material that blocked light transmission from the top view.
Collectively, the three proof-of-use experiments demonstrate various applications of GLUBscope to study  aquatic organisms.\\

\section{Conclusion}

We demonstrated GLUBscope, a modular DIY microscope 
made from 3D-printed parts and off-the-shelf components. GLUBscope was designed to enable imaging of samples from two orthogonal planes. This feature is useful for biomechanical investigation of aquatic organisms. Brightfield illumination was used for most of the study, and preliminary fluorescent imaging was also demonstrated. Better sources of illumination (brighter LEDs, or laser) similar to our prior efforts are needed  \cite{hedde2023spim} are needed to advance this feature. To record from two planes of imaging, the implementation required two cameras. We demonstrated the use of high-resolution/high-sensitivity and more economical options. This demonstration is valuable since the cameras are the most expensive component. As part of the report, proof-of-use experiments and analysis pipelines (specifically biomechanical studies of \emph{Hydra}) were demonstrated. Organisms in the current system were housed inside a cuvette (3.5 mL, 10 mm path) with four polished sides. In future experiments, microfluidics or other customized flow chambers, as described in \cite{hedde2017sidespim}, could be employed. It is important to note that investigating certain aspects of the hydrodynamics of freely moving aquatic organisms (e.g., sand-dollar larvae) may require modifications, larger volumes, and complex flow microenvironments. Among the existing DIY microscopes, the scale-free tracking system \cite{krishnamurthy2020scale,krishnamurthy2023active} is an elegant approach for providing a free-path for the movement of organisms in one direction. Conventional chambers, micro and millifluidics can still serve as valuable tools for many investigations. \\
GLUBscope is of interest in the biomechanics and hydrodynamics of aquatic organisms. Beyond their biological importance, these investigations could be crucial in mitigating threats due to climate change \cite{chan2012biomechanics}. Climate change threatens the planet's overall biodiversity. The recent Intergovernmental Panel on Climate Change (IPCC) reports specifically highlight a significant risk of permanent damage to aquatic organisms and their ecosystems \cite{IPCC2022}. Therefore, broader access to resources (e.g., microscopes) and closer investigation of the aquatic organisms and their responses to the emerging climate change challenges are needed.   

\newpage

\textbf {List of Figures}:
\begin{itemize}  
  \item Figure 1: GLUBscope - system overview.
  \item Figure 2: Build instructions part 1. hardware and optics. 
  \item Figure 3: Build instructions part 2: LEDs, electronics, and stage.
  \item Figure 4: System verification via two standard slides.
  \item Figure 5: GLUBscope - \emph{Hydra} case study.
\end{itemize}

\textbf {List of Supplementary Videos}:
\begin{itemize}  
  
  \item Supplementary Video 1: \emph{Hydra} side view.
  \item Supplementary Video 2: \emph{Hydra} top view.
  \item Supplementary Video 3: \emph{Hydra} side view 
  \item Supplementary Video 4: \emph{Hydra} top view - tracking. 
\end{itemize}

\textbf {List of Supplementary Figures}:
\begin{itemize}  
  \item Supplementary Figure -1: Preliminary results demonstrating fluorescent imaging. 
  \item Supplementary Figure -2:  GLUBscope -  case studies
\end{itemize}

\noindent
{\textbf{Animals and human rights:} Not applicable.}\\

\noindent
\textbf{CRediT author statement}\\

{\it 
\noindent
\textbf{Katherine Auer}: Methodology, Data Curation, Validation, Formal analysis, Investigation. \textbf{Brian Le}: Methodology, Data Curation, Validation, Formal analysis, Investigation. \textbf{David Lopez}: Methodology, Data Curation, Validation, Formal analysis, Investigation. \textbf{Justin P. Shum}: Methodology, Data Curation, Validation, Formal analysis, Investigation.  
\textbf{Brian Suarsana}: Methodology, Data Curation, Validation, Formal analysis, Investigation. \textbf{Ga-Young Kelly Suh}: Methodology, Validation, Resources. \textbf{Per Niklas Hedde}: Conceptualization, Methodology, Funding acquisition. \textbf{Siavash Ahrar}: Conceptualization, Methodology, Data Curation, Resources, Supervision, Project administration, Funding acquisition. All authors contributed to manuscript preparation.}\\

\noindent
\textbf{Declaration of Competing Interest}\\
\noindent
{\it The authors declare that they have no known competing financial interests or personal conflict or competing interest.}\\

\noindent
\textbf{Acknowledgements}\\
\noindent
\textit{The authors express gratitude to members of Ahrar-lab specifically Moses Villeda and Natalia Margaris, Prof. Rob Steele (UC Irvine), members of Pace-lab (CSULB), and CSULB Marine Lab specifically Yvette Ralph. }\\

\noindent
\textbf{Funding}\\
\textit{This work was supported in part by: 
1. CSUPERB award to SA. 2. National Institutes of General Medical Sciences (R21GM135493) award to PNH.\\ Granting agencies are not responsible for the content of the manuscript.}\\

\noindent
\bibliographystyle{unsrt}
\bibliography{References} 
\noindent

\begin{figure}[b]
\includegraphics[width=\textwidth]{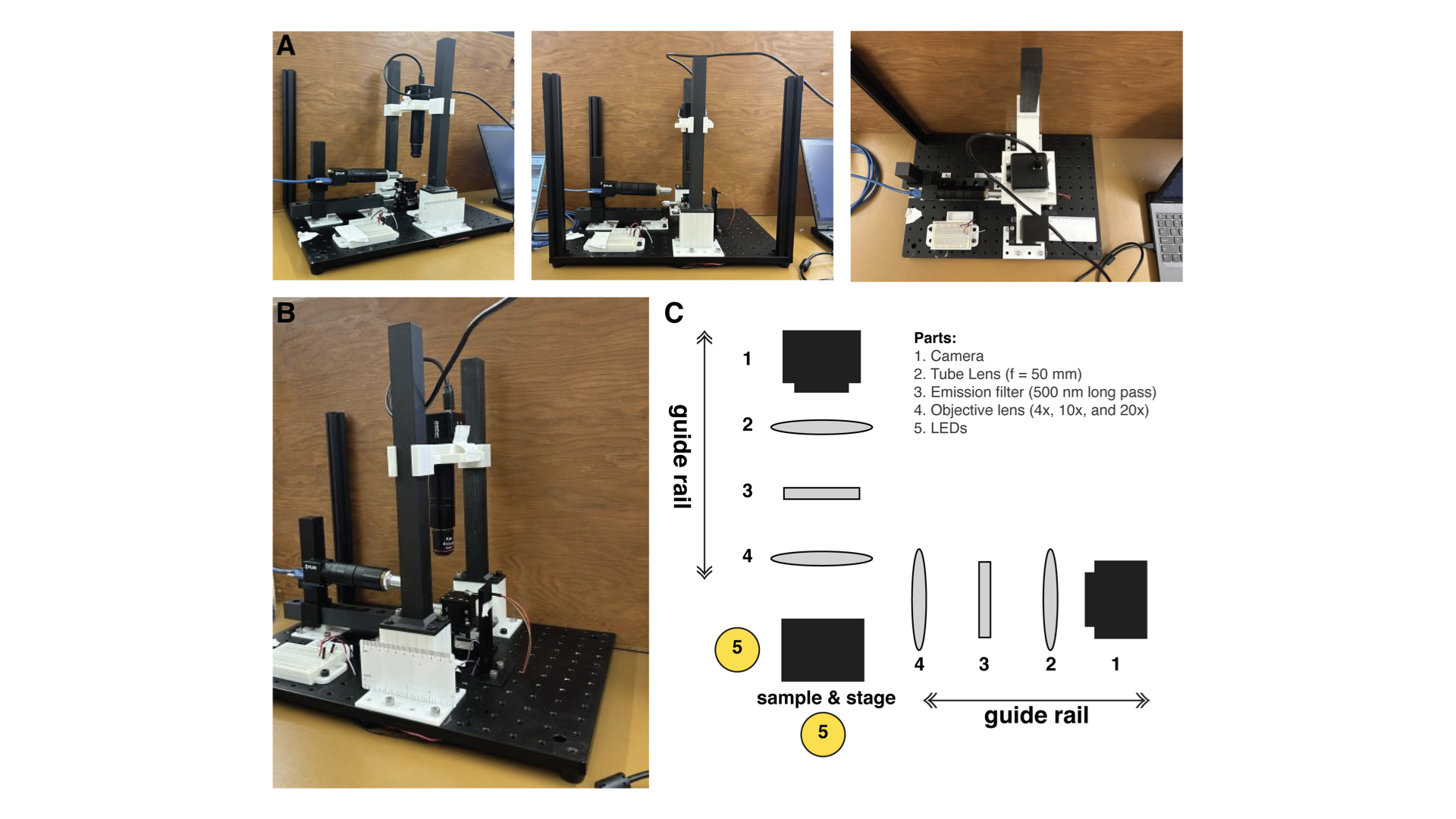}
\centering
\caption{\textbf {GLUBscope - system overview}. (A,B) Photographs of the finished GLUBscope. In this configuration two different cameras are used. (C) System block diagram of the major optical components for the GLUBscope.}
\end{figure}

\begin{figure}[b]
\includegraphics[width=\textwidth]{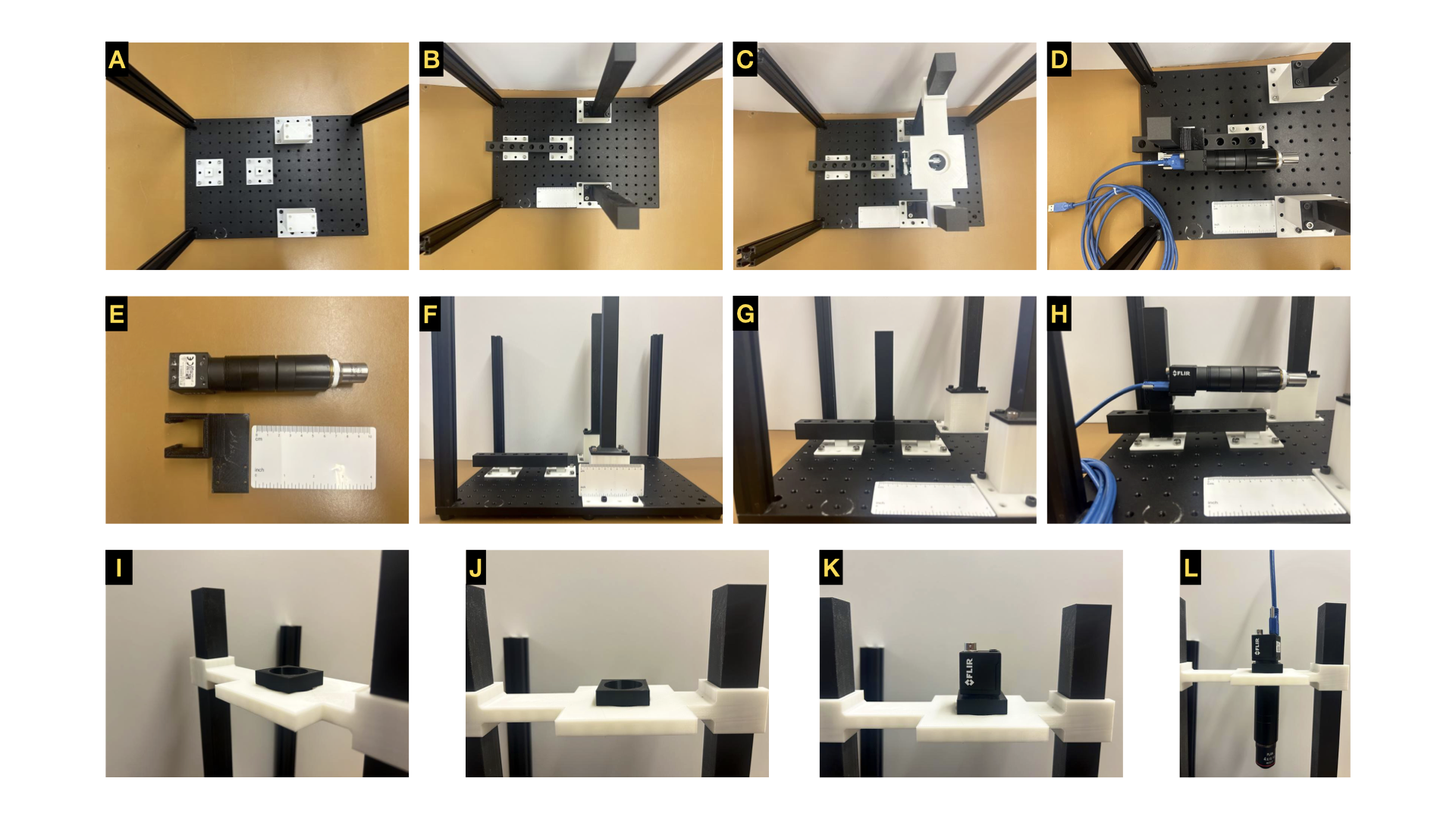}
\centering
\caption{\textbf {Build instructions part 1. hardware and optics. } (A:C)  Positioning of components (posts and guide rails) for imaging.  (D:H) Horizontal or side-view imaging components. (I:L) Vertical or top-view imaging components.}
\end{figure}

\begin{figure}[b]
\includegraphics[width=\textwidth]{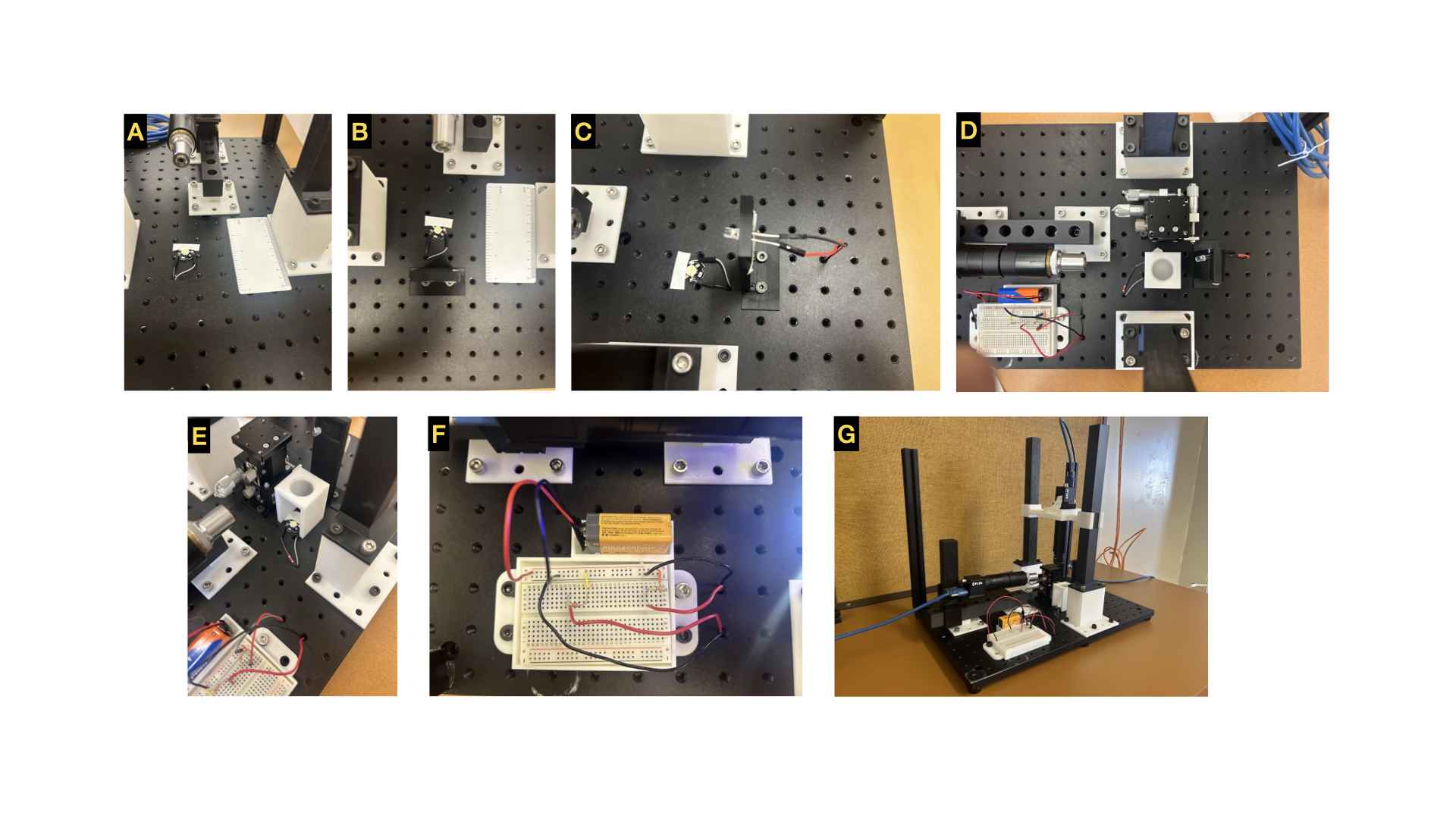}
\centering
\caption{\textbf {Build instructions part 2: LEDs, electronics, and stage}. (A:E) LEDs for vertical and side illuminations. Use of diffuser and holder are highligted in pannels D and E. In this example, bright field LEDs are demonstrated. (F) Electronic components to power the LEDs. (G) Fully assembled GLUBscope.}
\end{figure}

\newpage

\begin{figure}[b]
\includegraphics[width=\textwidth]{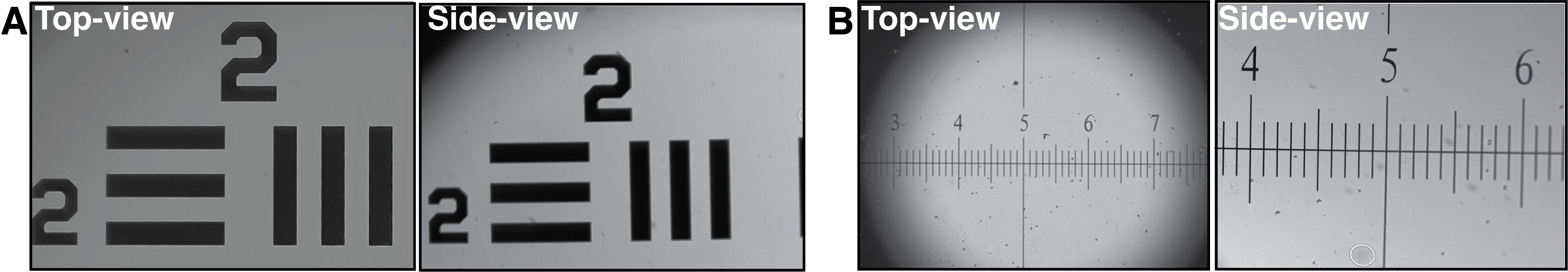}
\centering
\caption{\textbf {System verification via standard slides}. (A) Reference slide-1 from top and side view. (B) Reference slide-2 from top and side view}
\end{figure}

\begin{figure}[b]
\includegraphics[scale=0.87]{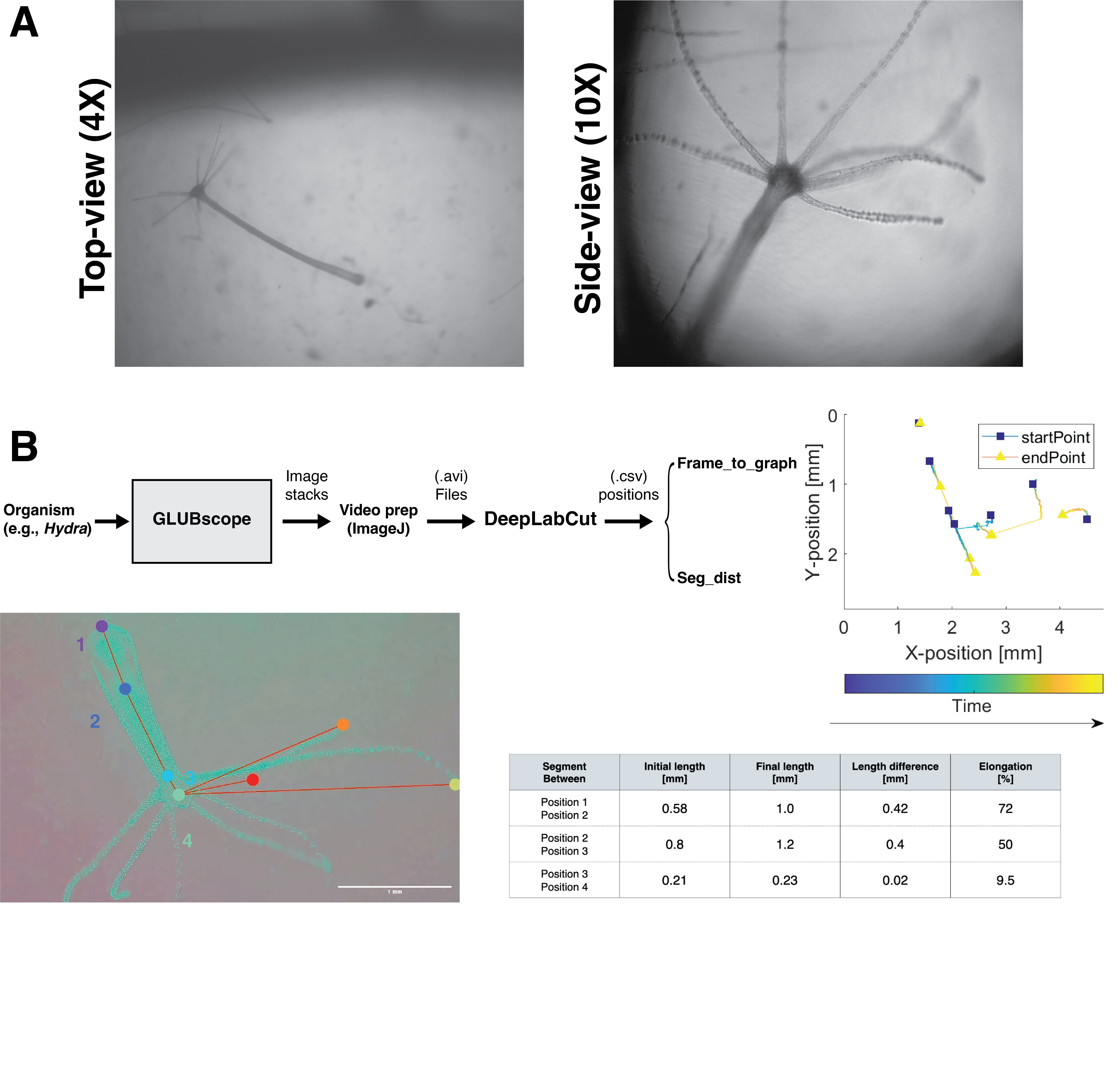}
\centering
\caption{\textbf {GLUBscope - \emph{Hydra} case study}. (A) \emph{Hydra} visualized from both top and side views. (B) Sample pipeline of tracking anatomical features (via DeepLabCut)and additional applications developed to estimate segment lengths and elongations. Plot demonstrates the staring and the final position of the anatomical features.
}
\end{figure}

\begin{figure}[b]
\includegraphics[width=\textwidth]{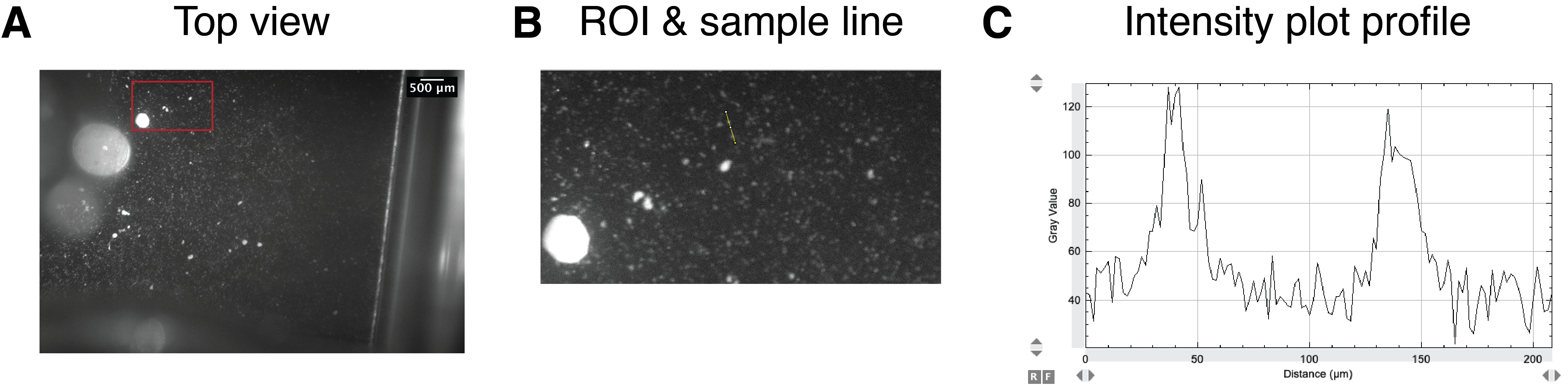}
\centering
\caption{\textbf {Supplementary Figure-1: fluorescent imaging.} (A) Green fluorescent beads (5 microns) were mixed in DIY water and vortexed inside a cuvette (3.5 mL volume and path length 1 cm). An off-the-shelf Green LED was used to illuminate the sample. (B) A region of interest (ROI) and a sample line for intensity measurements were selected. (C) The intensity profile along the sample line indicating the bead clusters. The signal to noise could be improved by using brighter excitation light source and reducing background .}
\end{figure}

\begin{figure}[b]
\includegraphics[scale=0.75]{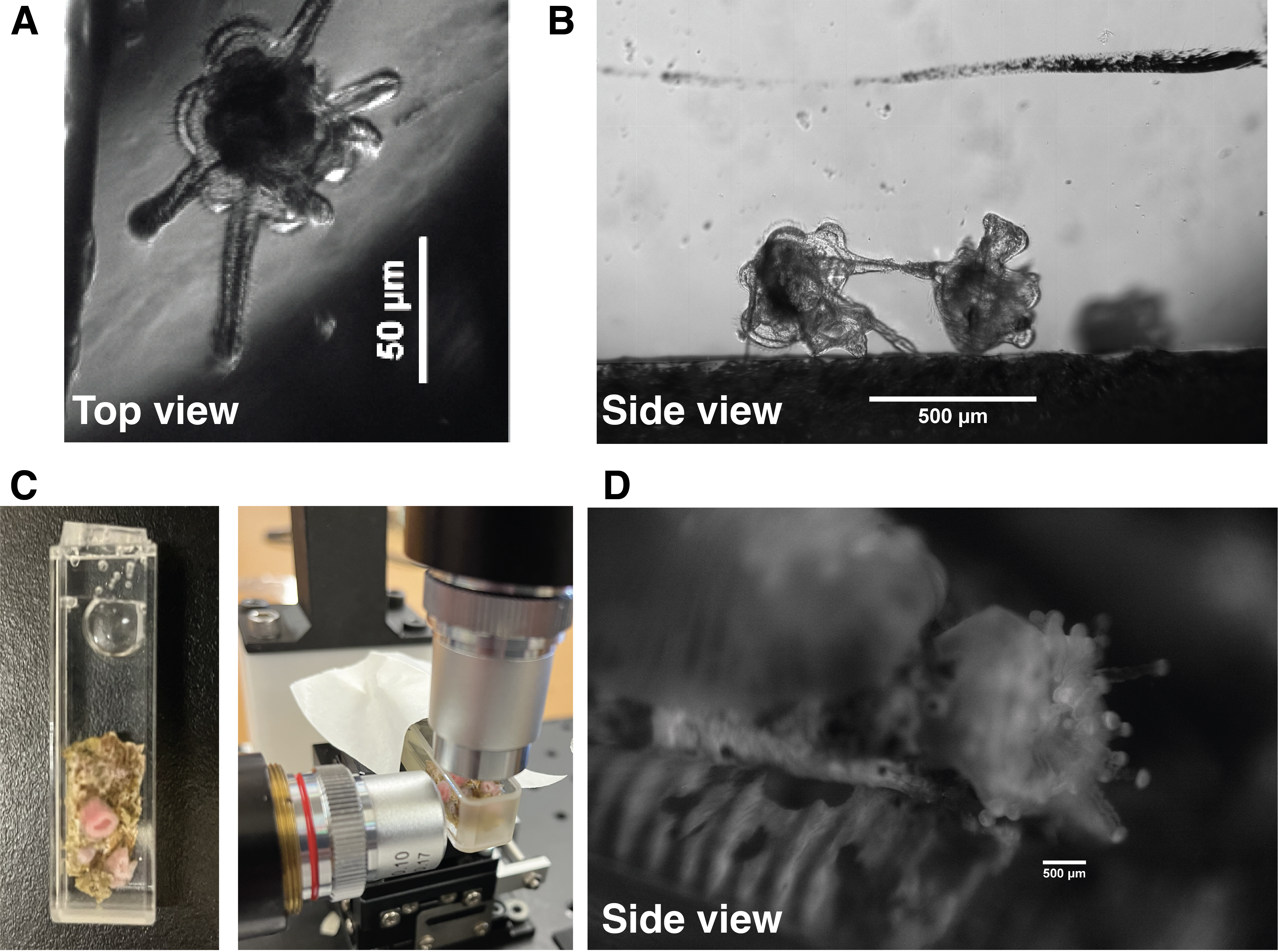}
\centering
\caption{\textbf {Supplementary Figure -2: GLUBscope -  case studies}. (A-B) top and side views of sand dollar larvae (C) Sea anemone attached to a rock inside the sample holder and on the stage. (D) Sea anemone imaged from the side view with GLUBscope.}
\end{figure}

\end{document}